\documentclass[letter]{aa}

\usepackage{graphicx}
\usepackage{txfonts}
\usepackage{lipsum}
\usepackage{lscape}             
\usepackage{placeins}           
\usepackage{comment}
\usepackage{xcolor}
\usepackage{esvect}

\usepackage{hyperref}
\hypersetup{
    colorlinks=true,
    linkcolor=black,
    urlcolor=black,
    citecolor=blue,
    }

\newcommand{\maglim}{\textsc{MagLim }}

\usepackage{acronym}
\newacro{GWTC-4.0}[GWTC-4.0]{fourth Gravitational-Wave Transient Catalog}    
\newacro{GW}[GW]{gravitational wave}
\newacro{LVK}[LVK]{LIGO-Virgo-KAGRA Collaboration}
\newacro{CBC}[CBC]{compact binary coalescence}
\newacro{PE}[PE]{parameter estimation}
\newacro{FAR}[FAR]{false alarm rate}
\newacro{BBH}[BBH]{binary black hole}
\newacro{EM}[EM]{electromagnetic}
\newacro{CMB}[CMB]{cosmic microwave background}
\newacro{MAP}[MAP]{maximum a posteriori}
\newacro{SNe Ia}[SNe Ia]{type Ia supernovae}
\newacro{APS}[APS]{angular power spectra}
\newacro{2PCF}[2PCF]{2-point correlation function}
\newacro{DES}[DES]{Dark Energy Survey}
\newacro{KiDS}[KiDS]{Kilo-Degree Survey}
\newacro{HSC}[HSC]{Hyper-Suprime Camera}
\newacro{LSST}[LSST]{Legacy Survey of Space and Time}
\acrodef{LIGO}[LIGO]{Laser Interferometer Gravitational-Wave Observatory}
\newacro{NF}[NF]{normalizing flows}
\newacro{IS}[IS]{importance sampling}

\begin{document}

\title{First measurement of the Hubble constant from a combined weak lensing and gravitational-wave standard siren analysis}

\titlerunning{First $H_0$ measurement from a joint weak lensing and standard siren analysis}


\author{Felipe Andrade-Oliveira\thanks{\email{felipe.andradeoliveira@physik.uzh.ch}}
        \and David Sanchez-Cid
        \and Danny Laghi
        \and Marcelle Soares-Santos
        }

\institute{Physik-Institut, Universität Zürich, Winterthurerstrasse 190, 8057 Zürich, Switzerland}

\date{Received January XX, 2026}

\abstract
    {We present a new measurement of the Hubble constant ($H_0$) resulting from the first joint analysis of standard sirens with weak gravitational lensing and galaxy clustering observables comprising three two-point correlation functions (3$\times$2pt). For the 3$\times$2pt component of the analysis, we use data from the Dark Energy Survey (DES) Year 3 release. For the standard sirens component, we use data from the Gravitational-Wave Transient Catalog 4.0 released by the LIGO-Virgo-KAGRA (LVK) Collaboration. For GW170817, the only standard siren for which extensive electromagnetic follow-up observations exist, we also use measurements of the host galaxy redshift and inclination angle estimates derived from observations of a superluminal jet from its remnant. Our joint analysis yields $H_0 = 67.9^{+4.4}_{-4.3}$~km~s$^{-1}$~Mpc$^{-1}$, a $6.4\%$ measurement, while improving the DES constraint on the total abundance of matter $\Omega_m$ by $22\%$. Removing the jet information degrades the $H_0$ precision to $9.9\%$. The measurement of $H_0$ remains a central problem in cosmology with a multitude of approaches being vigorously pursued in the community aiming to reconcile significantly discrepant measurements at the percent-level. In light of the impending new data releases from DES and LVK, and anticipating much more constraining power from 3$\times$2pt observables using newly commissioned survey instruments, we demonstrate that incorporating standard sirens into the cosmology framework of large cosmic surveys is a viable route towards that goal.}      


\maketitle
\nolinenumbers

\section{\label{sec:intro}Introduction}

The measurement of the present-day expansion rate, the Hubble constant ($H_0$), poses an empirical challenge to the $\Lambda$CDM cosmological model. $H_0$ values inferred from the early Universe through analyses of the \ac{CMB} data under the $\Lambda$CDM assumption~\citep{Planck:2018vyg, AtacamaCosmologyTelescope:2025blo, SPT-3G:2025bzu} are in significant ($\gtrsim5$-$\sigma$) tension with direct measurements obtained from late-Universe observables such as \ac{SNe Ia} calibrated using the cosmic distance ladder~\citep{Riess:2021jrx}. A series of analyses on both the early- \citep{Efstathiou:2019mdh, Hou:2017smn} and late-time  ~\citep{H0DN:2025lyy, DiValentino:2024yew} sides have been carried out to reconcile these measurements. While those efforts have achieved substantial progress in narrowing down the list of possible systematic effects, the problem persists. 

Motivated by this scenario, we present a new measurement of the Hubble constant resulting from an independent analysis of multi-messenger observables. Our approach builds upon the well-established multi-probe framework, known as the 3$\times$2pt analysis in the photometric survey community~\citep{DES:2021wwk, Heymans_2021, miyatake2023hypersuprimecamyear3}. This framework is mainly driven by the weak gravitational lensing, cosmic shear, \ac{2PCF} ~\citep{Secco_2022, Amon_2022, Wright_2025}, which is sensitive to the sum of baryonic and dark matter abundance ($\Omega_{m}$) and the amplitude of matter density fluctuations, combined with galaxy clustering \ac{2PCF} ~\citep{Rodr_guez_Monroy_2022}  and the cross-correlation of galaxy shear and galaxy position fields  ~\citep{Prat_2022, Pandey_2022, Porredon_2022}. We add to the 3$\times$2pt observables \ac{GW} distance indicators known as standard sirens~\citep{Holz:2005df}, which are binary coalescence events used for $\Lambda$CDM cosmological  measurements~\citep{LIGOScientific:2025jau} with redshift information gathered from \ac{EM} counterpart observations~\citep{LIGOScientific:2017adf}, catalogues of likely host galaxies~\citep{DES:2019ccw,LIGOScientific:2019zcs,Finke:2021aom,Palmese:2021mjm,Mastrogiovanni:2023emh,Bom:2024afj,Mukherjee:2022afz,deMatos:2025ifb}, and the mass spectrum of the binary population~\citep{LIGOScientific:2021aug,Karathanasis:2022rtr}. 
As information from follow-up observations of the EM counterpart of the standard siren GW170817 \citep{LIGOScientific:2017vwq,LIGOScientific:2017ync,Mooley:2018qfh} have been proven effective in enhancing its constraining power for cosmology \citep{Hotokezaka:2018dfi}, we also incorporate such information.  

Standard sirens and 3$\times$2pt are of particular interest to the $H_0$ problem because their observables are obtained without reliance on the cosmic distance ladder.  With this first combined measurement, we show that incorporating standard sirens into the multi-probe cosmology framework of large cosmic surveys is a viable route to constrain $H_0$ and thereby inform the ongoing debate.

\section{\label{sec:data}Data}

\subsection{\label{subsec:des}DES Y3}

The DES is a $grizY$ photometric survey that mapped 5000 deg$^2$ of the southern sky from Cerro Tololo, Chile, using the Dark Energy Camera (DECam,~\cite{Flaugher_2015}) between 2013 and 2019. The DES Y3 data release includes the first three years~\citep{Sevilla_Noarbe_2021}. 

We use the data vectors published in~\cite{DES:2021wwk} and available at the DES Data Management Portal\footnote{\href{https://des.ncsa.illinois.edu/releases/y3a2/Y3key-products}{https://des.ncsa.illinois.edu/releases/y3a2/Y3key-products}}. The weak lensing signal is extracted from a source galaxy sample drawn from a galaxy shape catalogue~\citep{Gatti_2021} comprising about 100 million galaxies. These are divided into four tomographic redshift bins extending to $z=2$, with a weighted effective source density of $n_{\rm eff}=5.9$ galaxies per arcmin$^2$ and a shape noise of $\sigma_{\rm e}=0.26$. Galaxy clustering is measured using the \maglim lens galaxy sample~\citep{Porredon_2021, Porredon_2022}, which contains 10.7 million galaxies split into six redshift bins. Redshift distributions for the source and lens samples are derived using self-organizing maps \citep{Giannini_2023} and the Directional Neighbourhood Fitting method \citep{De_Vicente_2016}, respectively.

\subsection{GWTC-4.0}
\label{subsec:gwtc}

The \ac{LVK} network of observatories~\citep{LIGOScientific:2025hdt} consists of two \acl{LIGO} \citep[\acsu{LIGO};][]{LIGOScientific:2014pky} detectors in the USA plus Virgo \citep{VIRGO:2014yos} in Europe and KAGRA~\citep{KAGRA:2020tym} in Japan. These four detectors are enhanced Michelson interferometers sensitive to gravitational waves in the $\sim$10-1000  Hz band~\citep{LIGOScientific:2025hdt}. The \ac{LVK} observing schedule is organized into observing runs during which one or more detectors are operational. The \acl{GWTC-4.0} \citep[\acsu{GWTC-4.0};][]{LIGOScientific:2025hdt,LIGOScientific:2025yae,LIGOScientific:2025slb} comprises \ac{GW} transient candidates accumulated between the first observing run (O1) and the end of the first part of the fourth observing run (O4a), which ended in January 2024. The catalogue reports the inferred source parameters under the hypothesis that these transients are caused by \acp{GW} emitted by \acp{CBC}. This catalogue has been used to study the population properties of \acp{CBC}~\citep{LIGOScientific:2025pvj} and to constrain late-time cosmological parameters~\citep{LIGOScientific:2025jau}.

We use data presented in~\citet{LIGOScientific:2025jau} and available in Zenodo\footnote{\href{https://doi.org/10.5281/zenodo.16919645}{https://doi.org/10.5281/zenodo.16919645}  (Version v1)}. We take the posterior samples for $H_0$ and $\Omega_m$ obtained using 142 \acp{CBC} out of the 218 \ac{GW} candidates included in \ac{GWTC-4.0}. We specifically choose the subset of the posteriors corresponding to the spectral sirens analysis, which relies exclusively on the GW data and uses the mass spectrum as the source of redshift information. GW170817  \citep{LIGOScientific:2017vwq,LIGOScientific:2017ync} is the only GWTC-4.0 bright siren (i.e., a \ac{CBC}  with a confirmed EM counterpart). To re-analyse GW170817 for this work, we also use the injection files~\citep{Essick:2025zed,LIGOScientific:2025yae} used to correct for \ac{GW} selection effects and the posterior samples for all its \ac{CBC} parameters, including the luminosity distance ($d_L$) and the source inclination angle ($\theta_{JN}$).

\subsection{GW170817 EM counterpart information}

Thanks to the discovery of its EM counterpart \citep{LIGOScientific:2017ync}, a wealth of multi-messenger data on GW170817 is available. This allows us to supplement the GW data with information derived from follow-up observations of the counterpart itself, as well as with archival data such as the spectroscopic redshift of the host galaxy, NGC 4993, and the velocity field at its position.

We use the host recession velocity $v=3327 \pm 72$~km~s$^{-1}$ \citep{Crook:2006sw} relative to the CMB and the radial peculiar velocity    $\langle v_p \rangle = 310$~$\pm$~150~km~s$^{-1}$ \citep{Carrick:2015xza}. We also use inclination angle constraint, $15\degr <\theta_{jet}< 29\degr$, obtained from observations of superluminal motion of the remnant's jet between 75 and 230 days post-merger~\citep{Mooley:2018qfh,Hotokezaka:2018dfi}.

\section{Methods}
\label{sec:method}

\subsection{Weak lensing and galaxy clustering}
\label{subsec:des-method}

In the 3$\times$2pt framework, cosmological information is inferred from overdensity and shear fields constructed from observations of galaxy positions and shapes. Field-level information is then compressed into the three \acp{2PCF}, shear-shear, galaxy-galaxy, and galaxy-shear, which in turn can be jointly compared to the cosmological model predictions \citep{krause2021darkenergysurveyyear, Friedrich_2021}. 

The theoretical framework is complemented with the modelling of known astrophysical and calibration systematic effects~\citep{krause2021darkenergysurveyyear}, such as galaxy bias~\citep{Porredon_2022}, intrinsic alignment~\citep{Blazek_2019}, lens magnification~\citep{Elvin_Poole_2023}, and redshift and galaxy shape calibration~\citep{DES:2020sjz, Giannini_2023, MacCrann_2021}. It also includes an analytical estimate of the signal covariance~\citep{Krause_2017, Fang_2020_fft}.

\subsection{Spectral sirens}
\label{subsec:spectral-method}

We use the results of the spectral siren analysis from the LVK \citep{LIGOScientific:2025jau}. Here we summarise their approach.

The \ac{GW} waveform produced by a \ac{CBC} with intrinsic mass $M$ at redshift $z$ is the same as that from an otherwise equivalent system of intrinsic mass $(1+z)M$ at redshift zero. This degeneracy between source-frame mass and redshift prevents us from directly obtaining the source redshift from the \ac{PE} of a \ac{CBC} transient.  However, redshift information can be obtained through the general framework of Bayesian hierarchical inference~\citep{Mandel:2018mve, Vitale:2020aaz}. This framework assumes that single-event \ac{CBC} parameters are drawn from a population distribution.  This distribution is described by a set of parameters that are inferred jointly with the cosmological parameters of interest and later marginalised over as nuisance parameters. 

\subsection{Bright siren}
\label{subsec:sirens-method}

To obtain multi-messenger cosmological posterior samples for GW170817, we start with the GW-only \ac{PE} samples from the \ac{LVK} analysis in~\citet{LIGOScientific:2018mvr} and incorporate the inclination angle information from \cite{Mooley:2018qfh} by running an importance sampling algorithm using the \ac{EM}-derived $\theta_{JN}$ range as an external flat prior. This procedure is similar to the one followed by \cite{Hotokezaka:2018dfi} and assumes that the jet emission is perpendicular to the orbital plane. 

The estimated $\theta_{JN}$ range is weakly dependent on $d_L$. We tested the impact of this dependency in our analysis and found it to be negligible.   

\subsection{\label{sec:inference}Cosmological inference}

To achieve our multi-messenger cosmology results, we combine the DES Y3 3$\times$2pt observables with 142 LVK standard sirens (141 spectral sirens plus the bright siren GW170817).   As both the DES and the LVK use Bayesian frameworks, in practise we combine posterior distributions, assuming they are independent. 

The choice to use LVK posterior samples from the spectral sirens method instead of the dark sirens method, which constructs line-of-sight redshift priors from a galaxy catalogue, avoids systematic uncertainties arising from covariances between the two types of probes, which would use overlapping galaxy catalogues.  The loss of constraining power due to this choice is minimal, as spectral sirens are much more constraining than dark sirens in the GWTC-4.0 LVK cosmology analysis~\citep{LIGOScientific:2025jau}.

In both the 3$\times$2pt and the standard sirens cases, we infer the cosmological parameters $\mathbf{s}_{\rm c} = \{H_0, \Omega_m\}$, while marginalising over a set of nuisance parameters $\mathbf{s}_{\rm n}$, which may include other cosmological parameters and astrophysical or calibration parameters. We adopted as priors $H_0 \in$ $\left[10,\,200 \right]$ km s$^{-1}$ Mpc$^{-1}$ and $\Omega_m \in [0, 1]$.

For the  $3 \times 2$pt analysis, we assume a Gaussian likelihood,
\begin{equation}
\label{ew:em_lk}
    \mathcal{L}\left(\mathbf{D}\,\big|\,\mathbf{s}_{\rm c}, \mathbf{s}_n\right) \propto \exp\left\{-\frac{1}{2}\mathbf{R}^\intercal \text{Cov}^{-1} \mathbf{R} \right\}\text{,}
\end{equation}
\noindent with $\mathbf{R} = \mathbf{D}-\mathbf{T}\left(\mathbf{s}_{\rm c}, \mathbf{s}_n\right),$ where $\mathbf{D}$ is the data vector collecting the \acp{2PCF} estimates, $\mathbf{T}(\mathbf{s}_{\rm c}, \mathbf{s}_n)$ is the theory prediction of the \acp{2PCF} for a given model described by the parameter sets $\mathbf{s}_{\rm c}$ and $\mathbf{s}_n$, and Cov is the covariance matrix of the three combined \ac{2PCF}. 
We refer to Table I of~\citet{DES:2021wwk} for the DES priors assumed in the inference. 
We introduce two changes with respect to the official DES Y3 analysis~\citep{DES:2021wwk}. First, we impose the above wider priors in $\Omega_m$ and $H_0$, 
to match the spectral sirens analysis. Second, we run the analysis with the \textsc{Nautilus}\footnote{Configuration: \texttt{n\_live}=10000, \texttt{n\_networks=16}}~\citep{lange2023nautilusboostingbayesianimportance} nested sampler algorithm within the \textsc{CosmoSis}\footnote{\href{https://cosmosis.readthedocs.io/}{https://cosmosis.readthedocs.io/}}~\citep{Zuntz:2014csq} software infrastructure. 

For the standard siren analysis, we assume a hierarchical Bayesian likelihood  given by: 
\begin{equation}
\label{eq:hierarchical_gw_lk}
    \mathcal{L}\left(D_{\rm GW}\,\big| \mathbf{s}_{\rm c}, \mathbf{s}_n \right) \propto \prod_{i}^{N_{\rm det}} \frac{\int \textup{d}\boldsymbol{\theta}\, \mathcal{L}(D_{{\rm GW},i}|\boldsymbol{\theta}) \, \pi(\boldsymbol{\theta} |\mathbf{s}_{\rm c}, \mathbf{s}_n)}{\xi(\mathbf{s}_{\rm c}, \mathbf{s}_n)} \text{,}
\end{equation}
\noindent where $D_{{\rm GW}}$ is the ensemble of \ac{GW} events data, $N_{\rm det}$ is the number of \ac{GW} detections used in the analysis,  $\mathcal{L}(D_{{\rm GW},i}|\boldsymbol{\theta})$ is the single-event \ac{GW} likelihood given the intrinsic single-event parameters $\boldsymbol{\theta}$ describing the binary, $\pi(\boldsymbol{\theta} |\mathbf{s}_{\rm c}, \mathbf{s}_n)$ is the population prior, and  $\xi(\mathbf{s}_{\rm c}, \mathbf{s}_n)$ is a term that represents the expected fraction of detected events in the population for a given set of cosmological and population parameters; this term accounts for \ac{GW} selection effects.
For the spectral siren analysis, we assume the population model \textsc{FullPop-4.0}~\citep{Fishbach:2020ryj,Farah:2021qom,Mali:2024wpq,LIGOScientific:2025jau} 
which includes the modelling of the merger rate and a mass model able to describe both neutron star and black hole mass ranges. We refer to Table 5 of~\citet{LIGOScientific:2025jau} for the \ac{LVK} priors assumed in the analysis.
For the spectral sirens, we use the existing outputs of the official LVK analysis. For GW170817, we also rerun the $H_0$ inference using the $d_L$ posterior samples with added jet information.
We perform our GW170817 re-analysis of $H_0$ with \texttt{gwcosmo}~\citep{Gray:2019ksv,Gray:2021sew,Gray:2023wgj}. 
The population prior is assumed to be a uniform distribution in mass ($[1,3]M_{\odot}$) with merger rate parameters fixed as in~\citet{LIGOScientific:2025jau} and with velocity relative to the Hubble flow $v_H = 3017 \pm 166$~km~s$^{-1}$ obtained by subtracting the recession velocity and peculiar velocity of NGC 4993~\citep{LIGOScientific:2017adf}. 
To facilitate combination later on and noting that GW170817 is at a very low-redshift ($\sim0.01$), and therefore is not sensitive to the $\Omega_m$ parameter, we augment the bright siren $H_0$ samples by adding a prior-dominated sample within the uniform interval $\Omega_m \in [0, 1]$.
We then combine the spectral and bright sirens posteriors to get the standard siren posterior distribution that we next combine with the $3 \times 2$pt posterior.

To evaluate the consistency of the posterior distributions from $3 \times 2$pt and standard sirens, we compute the distance between the medians of the $H_0$ posteriors, expressed in units of the combined standard deviation. We find a distance of 0.78$\sigma$, indicating good agreement between the two distributions. For $\Omega_{m}$, this requirement is automatically satisfied because neither spectral nor bright sirens place any significant constraints. Having verified that the consistency criterion is met, we proceed with combining the posteriors.

We combine the posterior samples from the $3 \times 2$pt and spectral sirens analyses using the normalising flow method presented in~\citet{gatti2024darkenergysurveyyear, raveri2024understandingposteriorprojectioneffects}, as implemented in \texttt{tensiometer}\footnote{\href{https://tensiometer.readthedocs.io/}{https://tensiometer.readthedocs.io/}}. This approach assumes that the individual posteriors are not mutually in tension, and that the priors on common parameters, $H_0$ and $\Omega_{\rm m}$, are consistent. 

\section{\label{sec:results}Results}

\begin{figure}[h]
    \centering
    \includegraphics[width=0.95\linewidth]
    {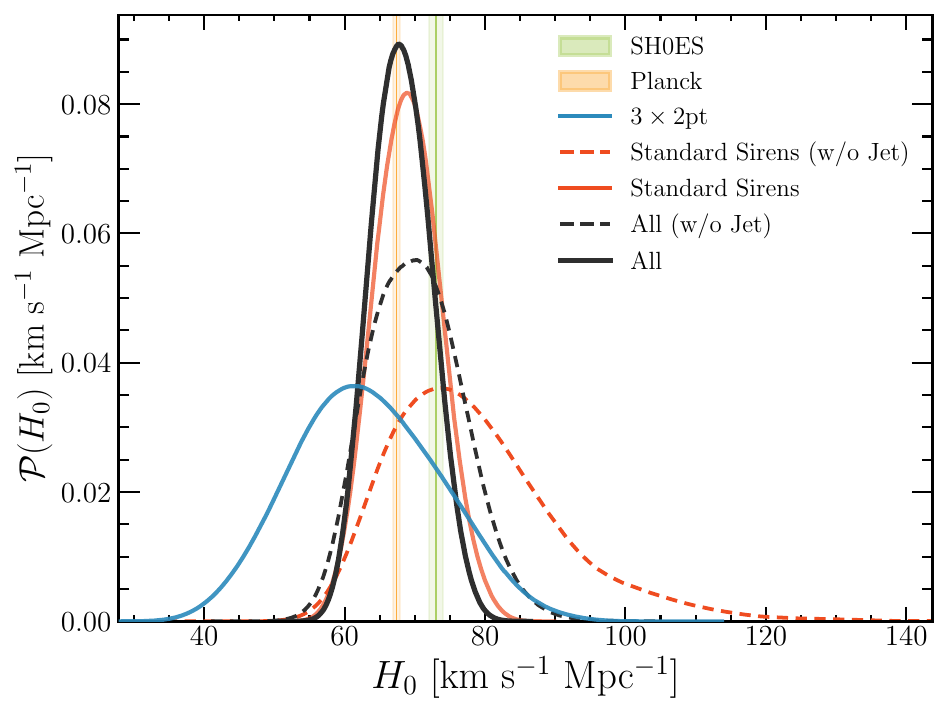}
    \caption{\label{fig:H0_posterior} Marginalised $H_0$ constraints from the $3 \times 2$pt analysis (blue), the spectral sirens plus GW170817 analysis (red), and their combination (black). Results obtained excluding the jet information are also shown (dashed lines). Orange and green bands are CMB~\citep{Planck:2018vyg} and SNe Ia~\citep{Riess:2021jrx} constraints, at 68\% credible interval.}
\end{figure}

Figure~\ref{fig:H0_posterior} shows our combined measurement as well as the contributions of each major component.  Table~\ref{tab:results} provides summary statistics including sub-components and alternative analysis choices. The $3 \times 2$pt-only yields $17\%$ precision constraint on $H_0$, while the standard sirens reach $7\%$. When combined, the resulting precision is $6.4\%$, with  $H_0=67.94^{+4.40}_{-4.34}$~km~s$^{-1}$~Mpc$^{-1}$ (median and $68\%$ credible interval). When the jet properties of GW170817 are excluded from the analysis, the precision in $H_0$ degrades to 15\% from standard sirens alone and 9.9\% after combination. The added constraining power of the 3$\times$2pt is 10-30\% depending on whether or not the jet information is considered. 
 
From Table~\ref{tab:results}, we also note that the contribution of GW170817 to the overall standard sirens analysis is significantly dominant only when the jet information is included. This is consistent with the fact that the constraining power of GW170817 without inclination angle information and the spectral sirens are similar for this dataset \citep{LIGOScientific:2025jau}. 
Also notable is the consistency between results with and without the jet information (within 1-$\sigma$). This indicates that potential biases due to mismodelling of the jet or misalignment between the jet-inferred angle and the true plane of the binary \citep{Muller:2024wzl} are subdominant in this analysis. 

\begin{table}[!b]
    \caption{\label{tab:results}
 $H_0$ and $\Omega_m$ constraints reported as the median and 68\% credible interval of the posterior probability distribution, accompanied by \ac{MAP} values in parenthesis. The top five rows correspond to the results shown in the figures. The bottom three rows are analysis variations shown for comparison.}
    \centering
    \resizebox{\columnwidth}{!}{%
    \begin{tabular}{l|c|c}
        \hline \hline
            & $H_0$ [km s$^{-1}$ Mpc$^{-1}$] & $\Omega_m$ \\ \hline\hline
        $3\times 2$pt & $62.80^{+11.49}_{-10.24}\;(60.02)$ & $0.352^{+0.055}_{-0.041}$ \\
        Standard Sirens (w/o Jet) & $76.45^{+13.12}_{-9.99} \;\; (76.78) $ & -- \\
        Standard Sirens & $68.95^{+4.75}_{-4.90}$ (68.21)  & -- \\
        All (w/o Jet) & $70.16^{+6.82}_{-7.02}\;\; (69.10)$ & $0.331^{+0.031}_{-0.031}$ \\
        All & $67.94^{+4.40}_{-4.34}$ (67.89) & $0.338^{+0.028}_{-0.028}$ \\
        \hline\hline
        $3\times 2$pt (w/ DES priors) & $65.18^{+9.94}_{-6.90}$ (58.03) & $0.345^{+0.036}_{-0.036}$ \\
        GW170817 (w/o Jet) & $78.47_{-11.98} ^{+25.87}\;\; (69.15)$
 & -- \\
        GW170817 &$68.77_{-4.75} ^{+4.94} \;\;(68.58)$
 & -- \\

 \hline \hline
    \end{tabular}
    }
\end{table}

\begin{figure}[h]
    \centering
    \includegraphics[width=0.95\linewidth]{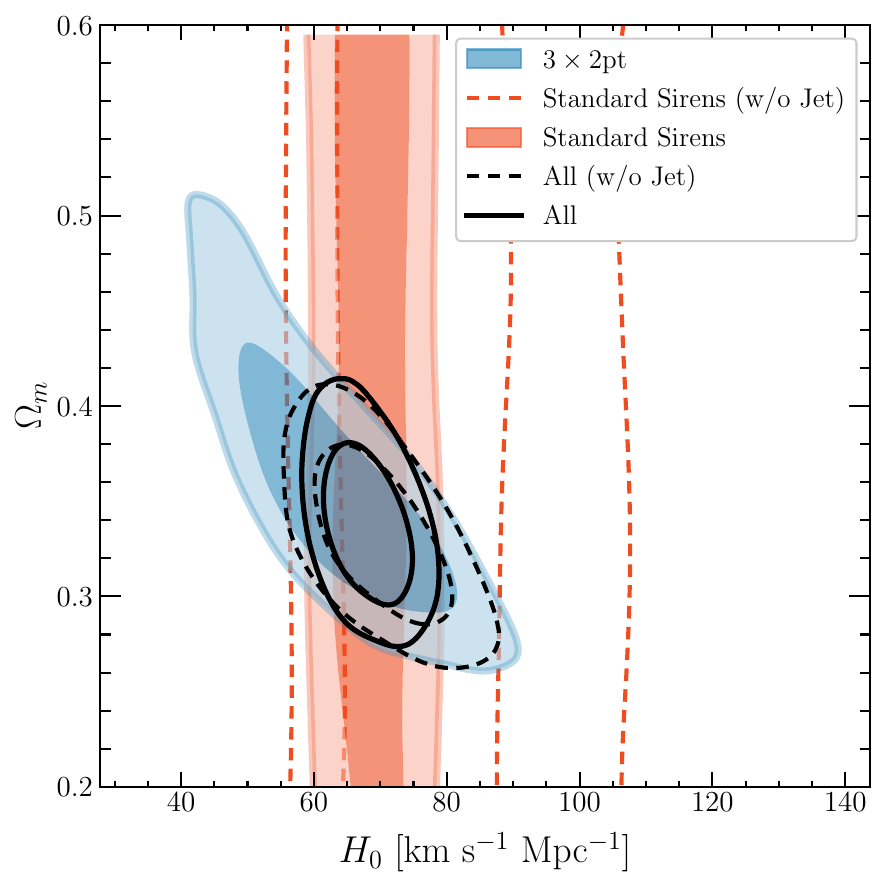}
    \caption{\label{fig:h0-om} Marginalised constraints on  $H_0$ and  $\Omega_{\rm m}$ from the $3 \times 2$pt (blue), the standard sirens (red) and the combination (black).  Results obtained without the GW170817 jet information are also shown (dashed). Contours show 1-$\sigma$ and 2-$\sigma$ credible levels.}
\end{figure}

Figure~\ref{fig:h0-om} shows our results in the $H_0$--$\Omega_m$ plane. 
Summary statistics for $\Omega_m$ are also provided in Table~\ref{tab:results}.
While the standard siren analysis presented here is not sensitive to $\Omega_m$ per se, its combination with 3$ \times $2pt has a significant impact in this plane. We gauge the gain in constraining power due to $3 \times 2$pt with standard sirens combination compared to the fiducial prior assumed in the DES Y3 analysis \citep{DES:2021wwk}, which is narrower in both $H_0$ (flat range [55, 91] km s$^{-1}$ Mpc$^{-1}$) and $\Omega_m$ (flat range [0.1, 0.9]). The precision in $H_0$ derived from the $3 \times 2$pt with narrower priors is 13\% (c.f., Table~\ref{tab:results}). 
Therefore, in comparison to the fiducial DES Y3 analysis (with tighter priors driven by external experiments), the information injected from the combination with standard siren doubles the precision in $H_0$ measurement.
Notably, the precision in the $\Omega_m$ measurement also benefits from the combination of $3 \times 2$pt and standard sirens, improving from 11\% to 8.2\% (a relative improvement of 22\%)  compared to the DES fiducial priors.

\section{\label{sec:conclusions}Conclusions} 

In this letter, we presented the first measurement of the Hubble constant from a combined probes analysis that integrates GW standard sirens to the 3$\times$2pt framework of joint weak lensing and galaxy clustering analyses from cosmic surveys. For this work, we used the DES Y3 source and lens galaxy catalogues and 142 GW events from the LVK GWTC-4.0 catalogue: 141 spectral sirens, for which redshift information is derived from the mass distribution of its population, plus one bright siren, GW170817, which has a well-studied \ac{EM} counterpart. 

We measured $H_0 = 67.9^{+4.4}_{-4.3}$~km~s$^{-1}$~Mpc$^{-1}$ ($6.4\%$ precision). We further showed that the precision degrades to $9.9\%$ when the jet information is removed from the analysis. Relative to the standard siren analysis alone, the 3$\times$2pt combination yields 10-30\% improvement on the $H_0$ uncertainties, depending on whether or not the jet information is used.
We also assessed the impact of our joint analysis on $\Omega_m$. While the standard sirens component alone is uninformative over this parameter, the combination leads to an improvement of 22\% in the $\Omega_m$ uncertainties, relative to the DES Y3 3$\times$2pt result alone.
This improvement demonstrates that standard sirens have sufficient constraining power to be informative when combined with $3 \times 2$pt probes, similarly to what is done with \ac{CMB}, \ac{SNe Ia}, or baryon acoustic oscillations. 

This result bodes well for future analyses using new data releases from DES and LVK, in the near term, as well as for the next generation of observatories in the long run. Full combined analyses, along the lines presented here, will leverage the combined constraining power of GW and 3$\times$2pt to reach percent-level precision on $H_0$ and advance the field of cosmology at a faster pace than otherwise expected.

\begin{acknowledgements} 
Funding for this work is provided by the University of Zurich (UZH) and the Swiss National Science Foundation (SNSF) under grant number 10002981. DL is supported by the UZH Postdoc Grant, grant no. [K-72341-01-01]. This material is based upon work supported by NSF’s LIGO Laboratory which is a major facility fully funded by the National Science Foundation. This project used public archival data from the Dark Energy Survey (DES). Funding for the DES Projects has been provided by the DOE and NSF(USA), MEC/MICINN/MINECO(Spain), STFC(UK), HEFCE(UK). NCSA(UIUC), KICP(U. Chicago), CCAPP(Ohio State), MIFPA(Texas A\&M), CNPQ, FAPERJ, FINEP (Brazil), DFG(Germany) and the DES Collaborating Institutions. Based in part on observations at Cerro Tololo Inter-American Observatory, NOIRLAB, which is operated by the Association of Universities for Research in Astronomy (AURA) under a cooperative agreement with the NSF.

\end{acknowledgements}

\bibliographystyle{aa}  
\bibliography{refs}

\end{document}